\documentstyle[11pt,paspconf,epsf]{article}

\begin{document}

\title{On the Formation of Bulges and Elliptical Galaxies in the Cosmological
       Context}
\author{V. Avila-Reese and C. Firmani}
\affil{Instituto de Astronom\'{\i}a, U.N.A.M., Apdo. Postal 70-264 \\
 04510 M\'exico D.F., M\'exico, and}
\affil{Astronomy Department, New Mexico State University \\
 P.O. Box 30001/Dept. 4500, Las Cruces, NM 88003-8001, USA}

\begin{abstract}
We study the formation of hot spheroidal systems within the frame of a scenario
where galaxy formation and evolution is related to the gentle mass aggregation
history and primordial angular momentum of protogalaxies, both defined by the
cosmological initial conditions. We explore two cases: (1) the hot spheroidal 
system forms from the dynamical instabilities of the stellar disks, and (2) 
the spheroidal systems are formed during the dissipative collapse of the gas 
before falling to the disk in centrifugal equilibrium. In the former case a 
good agreement with observations for late type galaxies is found. In the 
second case, contrary to recent claims, we find that the tidal stability 
criterion is not easily reached. The gas that dissipatively collapses within 
the dark matter halos should be very clumpy, and the clumps very dense, in 
order to avoid the tidal destruction of the star formation unities.
\end{abstract}

\keywords{galaxies: formation - galaxies: evolution - galaxies: ellipticals -
galaxies: spirals - star formation: theory - cosmology: theory}

\section{Introduction}

Galaxy formation and star formation (SF) are two related fundamental
problems of contemporary astrophysics and cosmology. Crucial questions on
these topics are: where, when, and how did stars form?; how the stellar
systems did evolve?; which is the origin of the Hubble sequence?. In the
present-day universe almost all stars are located either in rotationally
supported disks or in pressure supported spheroids. The stars could have
been formed into these systems through several mechanisms:

\textbf{(A) DISKS:} that stars form in disks is a more or less well understood
process, and actually is an ongoing observable phenomenon. The SF can be
induced in the thin gaseous disk by gravitational instabilities and it may
be self-regulated by an energy balance in the ISM along the vertical
structure of the disk (e.g., Firmani \& Tutukov 1992, 1994; Firmani,
Hern\'{a}ndez, \& Gallagher 1996; Firmani, Avila-Reese, \& Hern\'{a}ndez
1997). Once the stellar disks have been formed, there are two main ways to
partially or totally transform them into dynamically hot spheroidal systems: 
\textbf{(A1)} by internal dynamical instabilities (for example bars) in the
stellar disks, and \textbf{(A2)} by external interactions between disks. In
the former case it is expected the formation of bulge-like systems tightly
connected to the properties of the disks (the secular mechanism, e.g.,
Norman, Sellwood, \& Hassan 1996 and the references therein), while in the second case the
interaction, in particular if it is strong (merger), leads to the
destruction of the disk, and to the formation of a pressure supported
systems (e.g., Hernquist 1993). Nevertheless, these systems are not enough 
dense as the elliptical galaxies (EGs) are. \textbf{(A3)} If the dynamical
instabilities or the mergers occur when the disks are still plenty of gas,
then further gas concentration and subsequent bursts of SF are possible.
Dense hot spherical systems are expected to be formed in these cases.

\textbf{(B) HALO:} Disk systems form because the protogalaxies have some
initial angular momentum and the gas concentrates until the centrifugal
force equals the attractive gravitational force. However, it is possible
that the collapsing gas transform into stars before to reach the centrifugal
equilibrium. \textbf{(B1)} Gas thermal instabilities before or during the
dissipative collapse may produce population III stars (Lin \& Murray 1992)
and/or globular clusters (the stellar halo). \textbf{(B2)} The low angular
momentum gas clouds may attain high densities after a considerable
dissipative collapse in such a way that the SF is triggered; further the SF
can run by a self-regulated mechanism (e.g., Burkert 1994).

Since the galactic morphological type strongly depends on the environment,
it is also possible that physical mechanisms related to the environment like
the tidal stripping, the galaxy harassment, etc., are able to change the
morphology of galaxies.

The real understanding of the galaxy formation phenomenon can not be
accomplished without considering the cosmological context. The inflationary
CDM models together with the gravitational paradigm predict that cosmic
structures form through a hierarchical mass aggregation process. Within the
hierarchical clustering picture two general scenarios of galaxy formation
can be formulated: 1)\textbf{\ the merging scenario}, where the main
properties of galaxies are determined by the merger histories of
their dark matter (DM) halos (see Baugh 1998 in the present volume, and the
references therein), and 2) \textbf{the extended collapse scenario}, where
the properties of galaxies are mainly established by the combination of
three factors defined by the initial cosmological conditions: the mass,
the hierarchical mass aggregation history (MAH), and the primordial angular
momentum (Firmani et al. 1997, Avila-Reese 1998; Firmani \& Avila-Reese
1998; see also Gunn 1982, 1987; Ryden \& Gunn 1987). Here we shall explore
the problem of bulge and elliptical galaxy formation in the light of the
second scenario, for which the mechanisms (A1) and (B2) are adequate.

\section{Disk galaxies and the secular bulge formation mechanism}

In the extended collapse scenario the disks form inside-out with a gas
accretion rate dictated by the hierarchical mass aggregation process. The SF
is modeled as was described above (see point (A)). The secular mechanism of
bulge formation (the (A1) case) was followed through a simple physical
formulation: all the stars localized in the gravitational unstable regions
according to the Toomre criterion are transferred to a spherical component.
This formulation is in agreement with the results of detailed simulations
which show that the dynamical instabilities in the disks produce bars which
then dissolve forming dynamically hot regions (Norman et al. 1996 and the references therein).

Using the standard CDM model normalized to $\sigma _8=0.6$ as a
representative case, we obtain that as the model disks are redder and
with higher surface brightnesses, their gas fractions are lower and the
bulge-to-disk (b/d) ratios are larger, i.e. the models follow the
correlations of the Hubble sequence (Firmani \& Avila-Reese 1998). The b/d
ratios strongly correlates with the central surface brightnesses $\mu _{B_0}$
and do not correlate with the color indexes B-V (the models follow a
biparametrical sequence where $\mu _{B_0}$ and B-V are the two parameters).
The predicted b/d ratios for different masses, MAHs, spin parameters $%
\lambda ,$ and realistic cosmological models agree with the ratios derived
from the observations for disk galaxies (de Jong 1996). The b/d ratio is
mainly determined by $\lambda :$ more concentrated self-gravitating disks
imply more dynamical instabilities and therefore larger bulges. An
interesting prediction of the secular mechanism applied here is that the
more massive systems form their bulges earlier than the less massive ones
(Figure 1).

\begin{figure}
\vspace{6.0cm}
\includegraphics{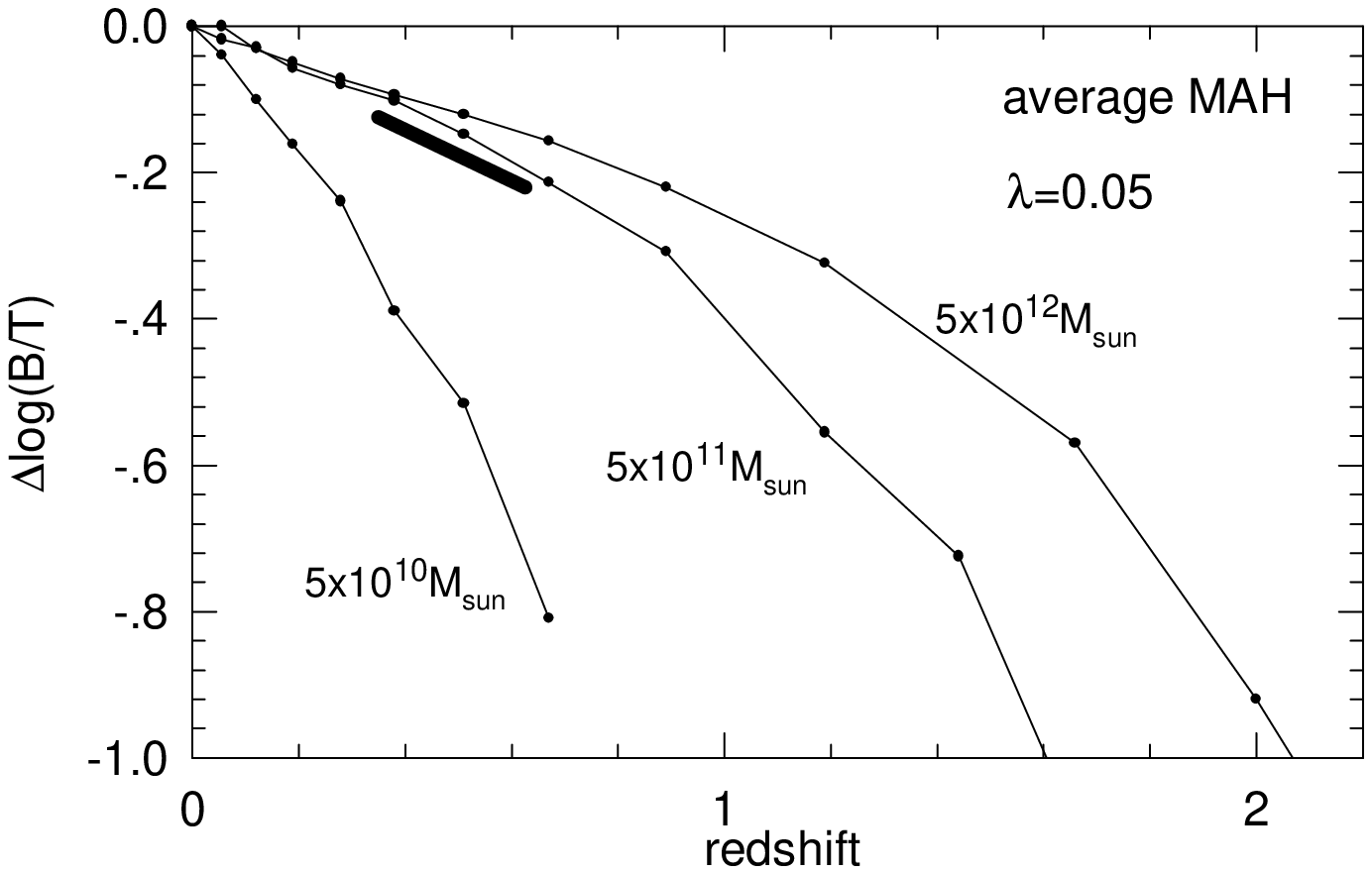}
\caption{Evolution of the bulge-to-total luminosity ratio for models with 
the average MAHs, $\lambda =0.05$ and for the SCDM, $\sigma_8=0.6$ model. 
$\Delta \log (b/t)\equiv \log (b/t)(z)-\log(b/t)(0).$ The thick segment 
corresponds to the slope inferred from the observational data presented in
Lilly et al. 1998. The more massive systems form their bulges earlier than 
the less massive ones.}
\end{figure}

For extreme cases (very low $\lambda 's$ or strong angular momentum transference)
the disks result very concentrated and completely self-gravitating (the (A3)
case). Although our models are not able to follow the dynamical evolution in
these particular cases, it is much probable that the disks will considerable
thicken or even completely destroy giving rise to hot spheroidal systems
whose properties could resemble those of the disky EGs. In this sense the
initial angular momentum and/or the ability of the protogalaxy to transfer
angular momentum (the last is related to the clumpyness, the environment and
the rapidity of the collapse of the protogalaxies) could be key factors
which determine the sequence of Disk-SO-Elliptical galaxies proposed for
example in Kormendy \& Bender (1996). It is important to emphasize that,
according to our models where the disks build-up within the evolving
cosmological DM halos and where the gravitational pull of the gas is
calculated, the circular velocities of the system become more and more
concentrated (steeply increasing, and after the maximum, decrease) as the
b/d ratio increases. In Figure 2 it is shown the rotation velocity of a
galactic system with a low spin parameter, $\lambda =0.03$ (thin solid line;
the b/d ratio in this case is 0.19). How are the gravitational potentials of
EGs? Rix et al. (1997) have intended to explore this difficult question and
they concluded that for the only EG they studied (NGC2434) the circular
velocity profile is flat. If this is confirmed for other EGs, then the
secular disk instability mechanism is not adequate to predict the formation
of EGs.

The simplified scheme of secular bulge formation applied to our disk galaxy
evolutionary models has shown to be predictive, at least for late type
galaxies. According to the secular mechanism, the properties (color,
scale lenghts, etc.) of the spheroids are closely associated to those of the
disks. For late type galaxies the observations indeed are in agreement with
these behaviors (Peletier \& Balcells 1996; de Jong 1996; Courteau et al.
1997). However, for early type galaxies ($<$Sb)
bulges and disks seem to have formed separately (e.g., Wyse, Gilmore, \&
Franx 1998). Therefore alternative scenarios should be explored in order to
explain early type galaxies.

\begin{figure}
\vspace{6.5cm}
\includegraphics{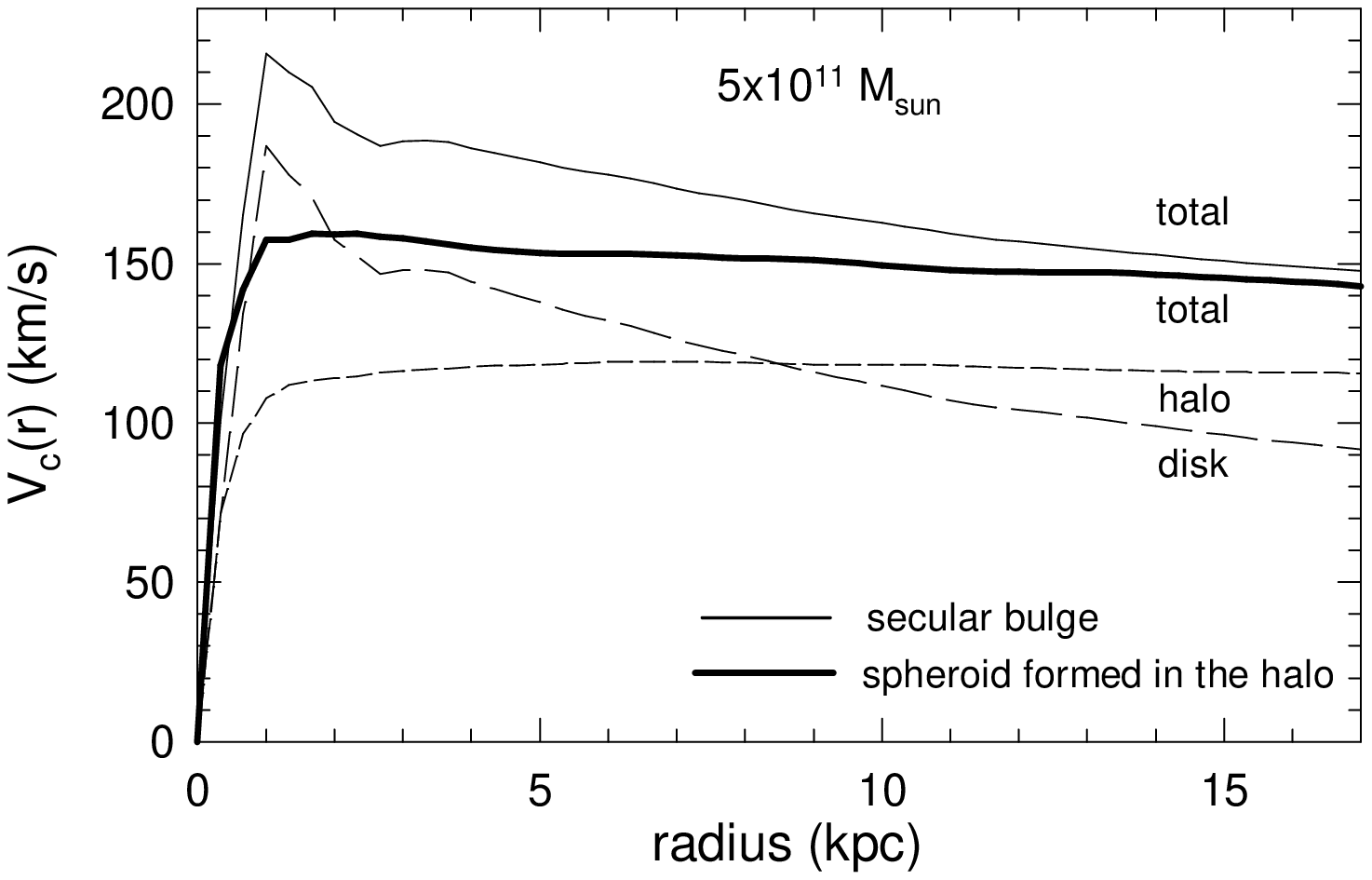}
\caption{Rotation curve decomposition of a galaxy model of 5x$10^{11}M_{\odot}$ 
with the average MAH and $\lambda =0.03$, and for the $\Lambda$CDM$_{0.3}$ 
model (thin lines). The solid line is the total rotation curve of the same 
model but when $62\%$ of the gas was transformed into stars in the halo 
before to fall to the disk.}
\end{figure}

\section{The dissipative collapse mechanism}
As was mentioned in the introduction the collapsing gas within the DM halos
can be transformed into stars before to reach the centrifugal equilibrium.
The first condition the gas should obey for this is the so called Roche
criterion: the gas clouds should be enough dense to avoid their destruction
by the global tidal forces. In a recent work Mao \& Mo (1998), using the DM
halo profiles predicted in the cosmological simulations (Navarro, Frenk, \&
White 1997) and the $\lambda -$distribution obtained in analytical and
numerical studies, calculated the baryon matter fraction that satisfies the
inequality

\begin{equation}
\rho _{gas}(r)>3\overline{\rho _{tot}}(<r)
\end{equation}
that roughly approximates the Roche criterion. Assuming that this gas
fraction is transformed into a spheroidal stellar system, Mao \& Mo
concluded that the predicted b/d ratios match the observations. We have
carried out a more detailed calculation which implies (i) the overall 
gravitational collapse and virialization of the primordial density 
fluctuations (the most probable cases are in agreement with the Navarro 
et al. halo  profiles; see Avila-Reese, Firmani, \& Hern\'{a}ndez 1998), 
(ii) the  dissipative collapse of the gas and its gravitational pull on 
the DM halo, and (iii) the formation of a rotationally supported disk 
within the evolving halo. We find that for the $\lambda -$distribution 
used by Mao \& Mo (1998) and for the realistic cosmological models, 
\textit{the tidal stability criterion (1) is almost never reached.} 
This discrepancy with the simplified analysis of Mo \& Mao is mainly due 
to the evolutionary effects and the gravitational pull of the gas over 
the DM halo, both considered in our case.

A more correct expression than (1) for the Roche criterion is:

\begin{equation}
\rho _{gas}(r)>2\overline{\rho _{tot}}(<r)*\left( 1-\frac 13\frac{d\lg M(r)}{%
d\lg r}\right) 
\end{equation}
In this case some fraction of the gas satisfies (2). For example, for the
$\Lambda$CDM$_{0.3}$, $\Omega _m=0.3,$ h=0.65 model and for the average MAHs 
this fraction goes from 4.0\% to 0.9 \% for $\lambda =0.03$ and $\lambda
=0.085,$ respectively. The range of variation of this fraction is much less
with the MAHs. Only for extreme cases of a very low $\lambda $ and a very
fast MAH, the gas fraction that satisfies (2) attain values of 10-20\%. In
conclusion our models show that for the \textit{uniformly} falling gas into
the center of the evolving cosmological DM halos the tidal stability Roche
criterion is not easily obeyed. To avoid the tidal destruction of the
SF unities the gas should have a clumpy distribution and much of the clumps
should be much denser than the average gas density. The dense clumps may
originate due to the non-homology of the collapse and/or due to the large
scale gas streams that collide in the center forming highly gas compressed
regions.

To obtain a considerable fraction ($>50\%$) of gas that satisfies
(2) the gas density should be incremented by factors larger than 20. In
these cases (2) is widely satisfied since large redshifts up to z$\approx 2-3
$ (for the $\Lambda$CDM$_{0.3}$ model). At these epochs the fraction of
collapsing gas into the DM halos that obeys the Roche criterion (2) is
0.8-1.0. If all this gas is transformed into stars then at z$\approx 2-3$
the bulge-to-total ratios of the model galaxies are near to one. In the
central regions of the galaxy clusters, where high density peaks complete 
their collapse early, these ratios may be maintained until the present epoch.

It is interesting to note that for these cases since the low angular 
momentum gas is not allowed to fall until its centrifugal equilibrium 
radius, but it is transformed into stars before, the resulting rotation 
curves are flatter than in the cases all the gas is allowed to fall into 
a disk in centrifugal equilibrium. In Figure 2 it is shown the rotation 
curve (thick solid line) of the same model corresponding to the thin line, 
but where 62\% of the gas has been transformed into stars in the halo.

\subsection{The star formation regime}

In analogy with the disk, one can expect that SF in the halo is
self-regulated by some mechanism once it was triggered. In the case of the
disk this self-regulation mechanism is commonly given by an energy balance in
the ISM (e.g., Firmani et al. 1996). The resulting SF timescale is $t_{SF}=%
\frac{2\varepsilon v_S}{v_g}t_{diss},$ where $\varepsilon $ and $v_S$ are
constants related to the negative feedback of the supernova energy
injection, $v_g$ is the gas velocity dispersion, and $t_{diss}$ is the time
scale of the gas (turbulent) energy dissipation (see Firmani et al. 1996). 
Since $v_g$ in the disks is relatively small, $\sim 6-10$ km/s, we obtain
typically  $t_{SF}\approx 100t_{diss}$, i.e. the negative feedback dominates
the SF timescales (inefficient SF). In the case of the halo the gas clouds
are infalling typically with velocities of $\sim 100-400$ km/s (the virial
velocities), therefore  $t_{SF}\approx 3-5t_{diss},$ i.e. the
negative feedback is not too important, the SF is more efficient than in the
disks. Since $t_{SF}$ is now near to $t_{diss}$, and $t_{diss},$ according
to our model results, depends on the galaxy mass (virial velocity), then $%
t_{SF}$ depends on mass. Thus, the b/d ratio will depend on mass. It is
probable that the physical conditions of the gas in the halo and in the disk
are quite different in such a way that a direct analogy is not possible. The understanding of the SF regime in the halo remains as an open crucial theoretical question.

\section{Conclusions }

The main conclusions drawn from our models of formation of hot spheroidal systems within the frame of the cosmological extended collapse
scenario are: (1) the \textit{secular mechanism of bulge formation }is able
to produce b/d ratios and correlations compatible with those observed in
late type galaxies; for very low angular momentum protogalaxies the whole disks
are dynamically unstable and S0/disky elliptical galaxies may form, however,
in these cases the circular velocities of the systems result strongly
decreasing. (2) for the \textit{dissipative collapse model of EG formation}
we find that the density of the gas that uniformly falls to the center of
the cosmological DM halos almost never is enough to obey the tidal stability
(Roche) criterion; a very clumpy gas dissipative collapse should be evoked
in order to form large hot spheroidal components. (3) If the SF regime in
the dissipative collapse model is self-regulated by an energy balance 
in the ISM analogous to that of the disk, then a strong dependence of the b/d
ratio on the galaxy mass is expected.

\end{document}